\documentclass[a4paper]{jpconf}
\usepackage{color,SIunits,bm}
\usepackage{pgfplots}
\usepackage{arydshln}
\usepackage{booktabs}
\usepackage{relsize}
\usepackage{dblfloatfix}
\usepackage{xcolor}
\let\origtau\tau
\renewcommand{\tau}{\scalebox{1.44}{$\origtau$}}
\usepgfplotslibrary{colormaps}
\pgfplotsset{compat=1.5}

\usepackage{graphicx}
\usepackage{caption}

\usepackage{caption}
\usepackage[left=2.6cm, right=2.6cm, top=4cm]{geometry}

\usepackage{amsmath}
\usepackage{booktabs}
\usepackage{graphicx}

\usepackage{multicol}
\usepackage{wrapfig}

\def\bite{\rm Bi_2Te_3}
\def\bise{\rm Bi_2Se_3}
\def\sige{\rm SiGe}
\def\pbte{\rm PbTe}

\begin{document}
\twocolumn[
\begin{@twocolumnfalse}
\title{Non-universal Scaling of Thermoelectric Efficiency in 3D and 2D Thermoelectric Semiconductors}

\author{Kevin Octavian$^{1,a)}$, Eddwi H. Hasdeo$^{2,3, b)}$}

\address{$^1$Engineering Physics, Faculty of Industrial Technology, Institut Teknologi Bandung, Indonesia}
\address{$^2$Research Center for Physics, Indonesian Institute of Sciences,
  South Tangerang, Indonesia}
\address{$^3$Department of Physics and Material Science, University of Luxembourg, Luxembourg}
\ead{$^{a)}$kevinn.octavian@s.itb.ac.id\\
$^{b)}$eddw001@lipi.go.id}

\date{-}
\begin{abstract}
We performed the first-principles calculation on common thermoelectric
semiconductors $\rm Bi_2Te_3$, $\rm Bi_2Se_3$, $\rm SiGe$, and $\rm
PbTe$ in bulk three-dimension (3D) and two-dimension (2D). We found that miniaturization of materials does not generally increase the thermoelectric figure of merit ($ZT$) according to the Hicks and
Dresselhaus (HD) theory. For example, $ZT$ values of 2D $ \rm
PbTe$ (0.32) and 2D $ \rm SiGe$ (0.04) are smaller than their 3D counterparts (0.49 and 0.09, respectively). Meanwhile, the $ZT$ values of 2D
$\rm Bi_2Te_3$ (0.57) and 2D $\rm Bi_2Se_3$ (0.43) are larger than the bulks (0.54 and 0.18, respectively), which agree with HD theory. 
The HD theory breakdown occurs because the band gap and 
band flatness of the materials change upon
dimensional reduction. 
We found that flat bands give a larger electrical conductivity ($\sigma$) and electronic thermal conductivity ($\kappa_{el}$) in 3D materials, and smaller values in 2D materials.
In all cases, maximum $ZT$ values increase proportionally with the band gap and saturate for the band gap above $10\ k_BT$. 
The 2D $\bite$ and $\bise$ obtain a higher $ZT$ due to the flat corrugated bands and narrow peaks in their DOS. Meanwhile, the 2D PbTe violates HD theory due to the flatter bands it exhibits,
while 2D SiGe possesses a small gap Dirac-cone band.  \\

\end{abstract}

\end{@twocolumnfalse}
]

\section{Introduction}

Thermoelectric (TE) materials are useful for generating electricity from waste heat without any
moving parts. Despite decades of research in this field, TE efficiency remains low and stagnant
at 10\%. This efficiency corresponds to a dimensionless figure of merit ($ZT$) that equals to
unity, which is defined as
\begin{equation}
ZT = \frac{S^2 \sigma}{\kappa_{el} + \kappa_{ph}} T,
\end{equation}
where $S$ is the Seebeck coefficient, $\sigma$ is the electric conductivity, $\kappa_{el}$ is the electronic
thermal conductivity, $\kappa_{ph}$ is the phonon thermal conductivity, and $T$ is the effective temperature. High electric conductivity is needed
to obtain a high $ZT$ value, but increasing it also increases the thermal conductivity, following
the Wiedemann-Franz law $\kappa_{el} = \sigma L T$, where $L$ is the Lorenz number. This
proportionality reduces $ZT$ value. It is hard to achieve a $ZT$ value over unity due to this
relation.

Another factor that reduces the $ZT$ further is the fact that
metals exhibit a low Seebeck coefficient, while insulators have the opposite characteristics. 
Thus good
TE materials usually come from semiconductor materials. There exists a range of band gaps \cite{Hasdeo2019, Mahan1994} and band widths \cite{Chen2011} which
give the optimal $ZT$ value.

One way to push $ZT$ value beyond unity is through the miniaturization of
materials as initially proposed by Hicks and Dresselhaus (HD)
\cite{Hicks1993, Hicks1993_1D}. The density of states in 2D and 1D materials show sharp
steps and the Van Hove singularities,
respectively, which are responsible for the increase of the Seebeck
coefficient and hence the $ZT$ value as well. The breakthrough of $ZT$
values has been observed in 1D and 2D nanostructured materials, such
as hierarchical PbTe \cite{Biswas2012}, silicon nanowires
\cite{Hochbaum2008}, nanostructured BiSbTe \cite{Poudel2008}. However,
the enhancement due to miniaturization of materials only works when
the confinement length is smaller than its thermal de Broglie
wavelength \cite{Hung2016}. With recent advances in crystal
growing of 2D materials, it is possible to have one or few
atoms-thick 2D materials that satisfy small confinement lengths.

Moreover, HD theory simply assumes parabolic bands that retain the same band gaps and band flatness as the dimension
changes. In reality, these quantities strongly depend on the geometry
and the dimension of the materials, and as a result, they will affect
TE transport. In this paper, we investigate several common semiconductors TE to check the limitation of the HD theory.

We investigate the 3D and the 2D structures of $\bite$, $\bise$,
$\pbte$ and $\sige$. The TE properties were calculated by using the
first-principles calculation and the Boltzmann transport equation. Additionally, these results can be compared with a  simple two-band model to understand the dependence of TE properties on dimensionality, band gap, and band flatness. While the bulk
states of these materials are considered as good TE materials, the TE
properties of the 2D structures remain in early-stage research. The single
quintuple layer (QL) of $\bite$ and $\bise$ have been experimentally
fabricated through exfoliation \cite{Teweldebrhan2010, Sun2012}. 
In the recent
study \cite{Liu2015}, it has been shown that the (001) PbTe monolayer
turns into a 2D topological crystalline insulator while SiGe has a
graphene-like structure on its 2D surface (siligene) \cite{Jamdagni2015}.

Our results show
that dimension reduction changes the band gaps and the
flatness of the band. From the analysis of the two-band model, we show that the band flatness keeps the Seebeck coefficient intact and reduces the $\sigma$ and $\kappa_{el}$ values in 2D materials,
while in 3D materials, it increases  $\sigma$ and $\kappa_{el}$ due to the different density of states. Overall, the maximum $ZT$ values increase proportionally with band gap and saturate when band gap above $10\ k_B T$ in both the 2D and 3D materials. 
As a result, 2D PbTe, which exhibits relatively flat bands, and 2D SiGe, which has a low band gap, 
have a low $ZT$ value, and violate HD theory. On the other hand, $\bite$ and $\bise$ agrees with HD theory 
because of their flat corrugated bands~\cite{Mori2016} and a lot of narrow peaks on the DOS giving an enhancement in their $ZT$ values.
Our results also show that the $ZT$ values of the investigated
materials increase as the temperature increase, except for $\bite$.

\section{Methods} 
We used Quantum Espresso \cite{Giannozzi2009} to perform all density functional theory (DFT)
calculations with the projected augmented wave (PAW) method \cite{Kresse1999}. The
generalized-gradient approximation (GGA) of Perdew-Burke-Ernzerhof (PBE) functional was used as
the exchange-correlation \cite{Perdew1992}. The plane wave's cutoff energy and the charge density
were set to 60 Ry and 720 Ry, respectively. The Monkhorst-Pack scheme \cite{Monkhorst1976} was
used to integrate the Brillouin zone in the self-consistent calculations with a k-point mesh of
10 x 10 x 10 for the bulk materials and 10 x 10 x 1 
\begin{table}[b]
    \centering
    \caption{Fitted relaxation times and phonon thermal conductivities at 300K}
    \scalebox{0.8}{
    \begin{tabular}{c c c }
    \toprule
        Material & $\tau\ (10^{-14}\ s)$ & $\kappa_{ph} (W/mK)$ \\
    \midrule
        $\bite$ & 2.8 & 1.37 \\
        $\bise$ & 0.7 & 1.00 \\
        $\pbte$ & 1.1 & 2.15 \\
        $\sige$ & 0.8 & 4.60 \\
    \bottomrule
    \end{tabular}}
    \label{tab.tau}
\end{table}
for the 2D materials. A vacuum layer of 35 $\textrm{\r{A}}$ is used for the 2D calculations. 
The convergence criteria for structure optimization was taken to be less than $10^{-3}$ eV and
less than $0.025$  eV $\rm \r{ A}^{-1}$ for the total energy and the total force, respectively. 

\begin{figure}[t]
    \centering
    \includegraphics[scale=0.12]{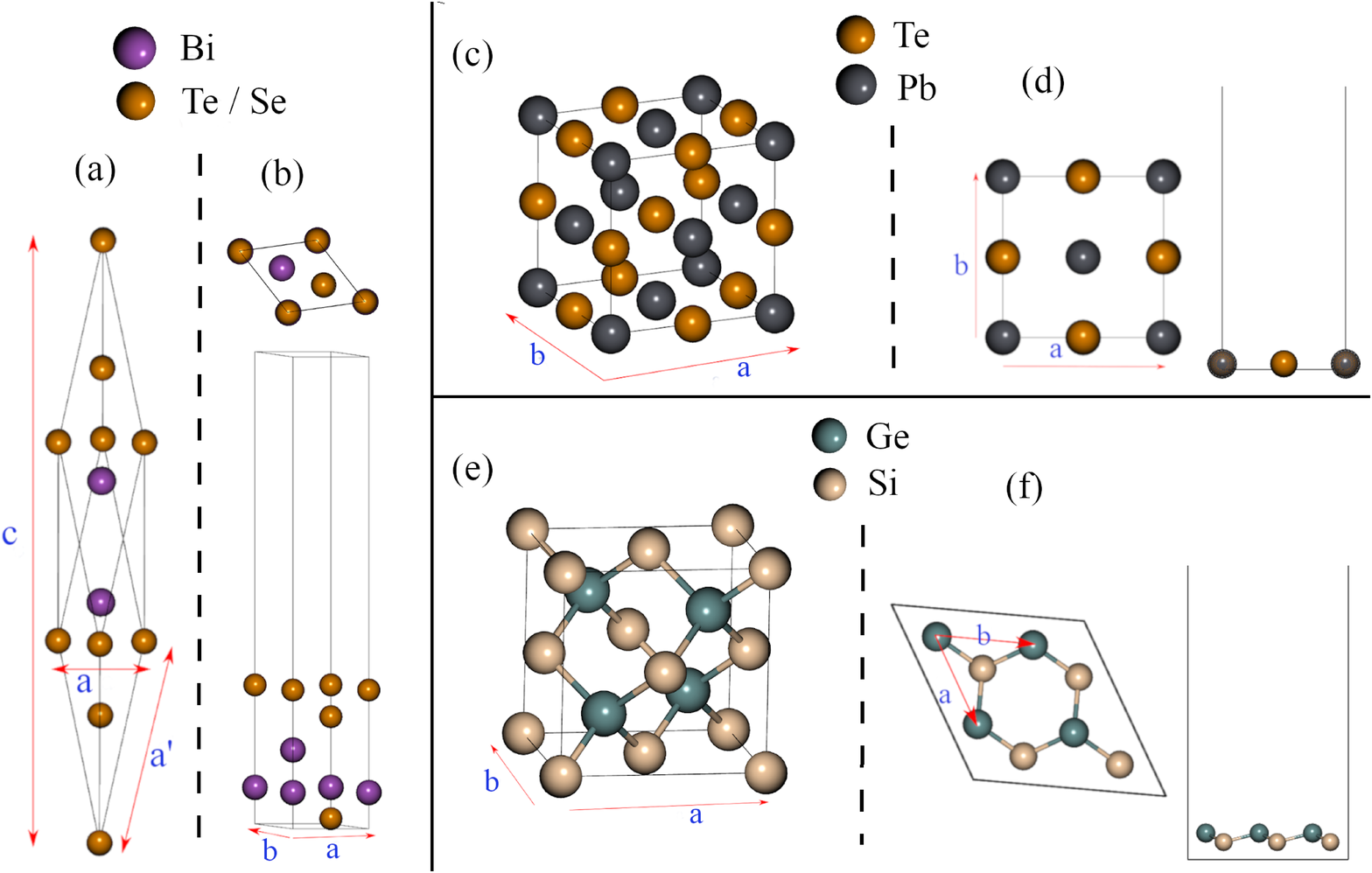}
    \caption{Crystal structures of (a) bulk $\bite$ or $\bise$, (b) a single QL of $\bite$ and $\bise$, (c) bulk PbTe, (d) PbTe(001) monolayer, (e) bulk SiGe, and (f) SiGe(001) monolayer}
    \label{fig.structure}
\end{figure}

The semi-classical Boltzmann equations encoded in the BoltzTraP program
was used to evaluate the transport properties \cite{Madsen2006}. To
give a better result, a denser k-mesh of 40 x 40 x 40 and 80 x 80 x 1
were used for the bulk materials and the 2D materials,
respectively. Relaxation time ($\tau$) and phonon thermal conductivity
($\kappa_{ph}$) were required to evaluate the dimensionless figure of merit ($ZT$) of a
material. The values presented in Table~\ref{tab.tau} are obtained by
using the method described in Appendix A of Supplementary Material \cite{supplementary} for all bulk materials. We employed the
same values to the corresponding 2D materials.

\section{Results and Discussion}
\subsection{Structural Properties}
The phase groups of the materials that we used are as follow, \textit{R3m} for $\bite$ and
$\bise$, \textit{Fm3m} for $\pbte$, and \textit{F43m} for $\sige$. As for the 2-dimensional
structures, the quintuple layer (QL) of $\bite$ and $\bise$, and (001) surface layer of $\pbte$
and $\sige$ were used. In this study we only investigated a single layer of each materials. All
of the structures are presented in Fig.~\ref{fig.structure}.

\begin{figure*}[t]
    \centering
    \includegraphics[scale=0.5]{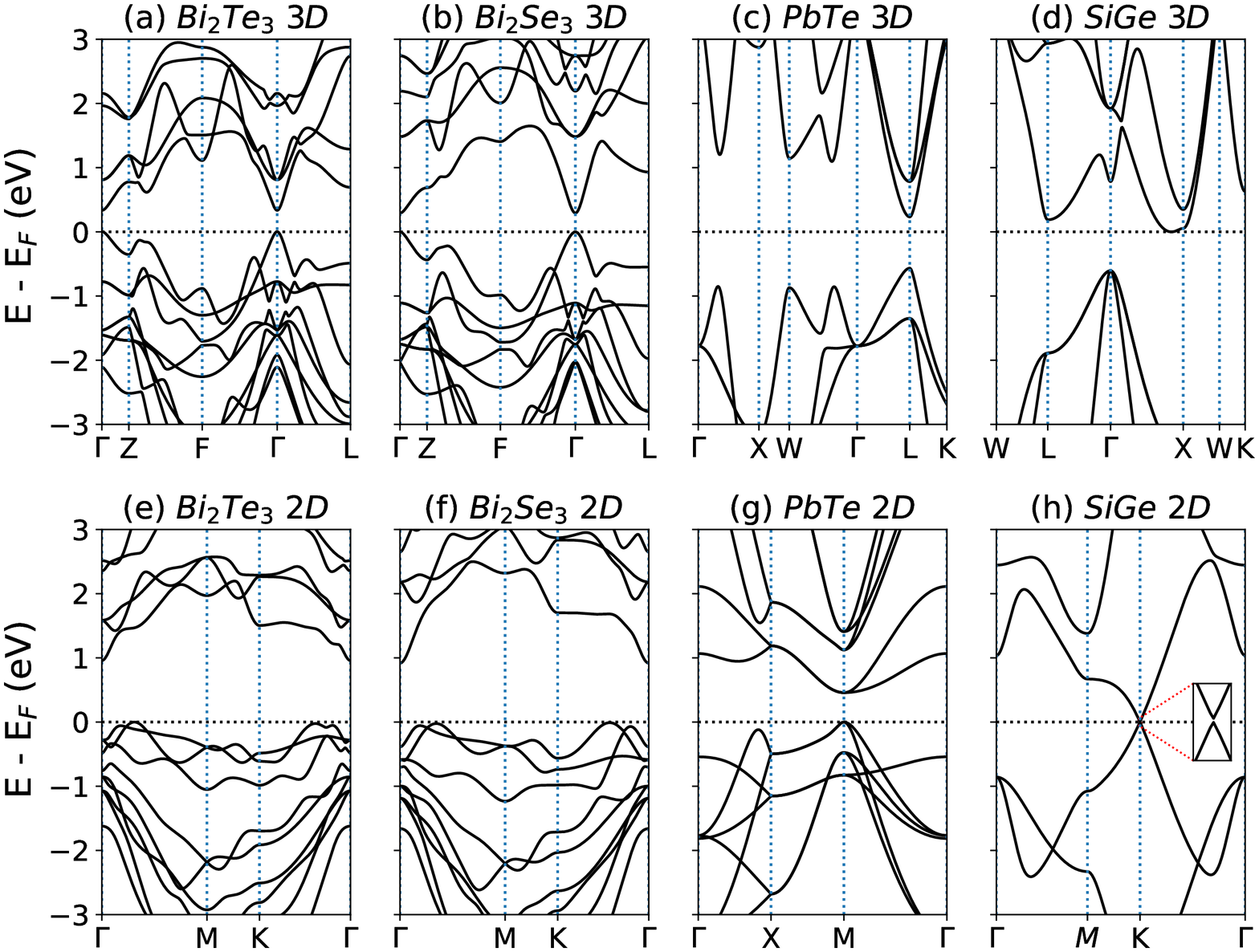}
    \caption{The calculated band structure of (a) bulk $\bite$, (b) bulk $\bise$, (c) bulk PbTe,
    (d) bulk SiGe, (e) single QL of $\bite$, (f) single QL of $\bise$, (g) PbTe monolayer, and
    (h) SiGe monolayer}
    \label{fig.band}
\end{figure*}

The results of our structure optimization are shown in Table. 3 of Supplementary
Material. The error between our results and the experiment values are less than 2 \%. A further
reduction in error could be obtained by using tighter convergence criteria. Bulk $\bite$ and
$\bise$ both have the same hexagonal close-packed (HCP) crystal structure. Bulk PbTe has a NaCl face-centered cubic crystal structure. The surface of PbTe
in (001) direction possesses a similar lattice constant with the bulk structure, although it is
stated in \cite{Jia2017} that the lattice constant of (001) few-layers decreases drastically, but
the magnitude is unclear for the monolayer PbTe. The (001) surface of SiGe (siligene) has a similar
structure with graphene. Nevertheless, unlike planar graphene, siligene possesses a buckling
structure. The calculated buckling amplitude is 0.58 $\textrm{\r{A}}$, which agrees with the
previous theoretical work \cite{Sannyal2019}.

\begin{figure}[t]
    \centering
    \includegraphics[scale=0.4]{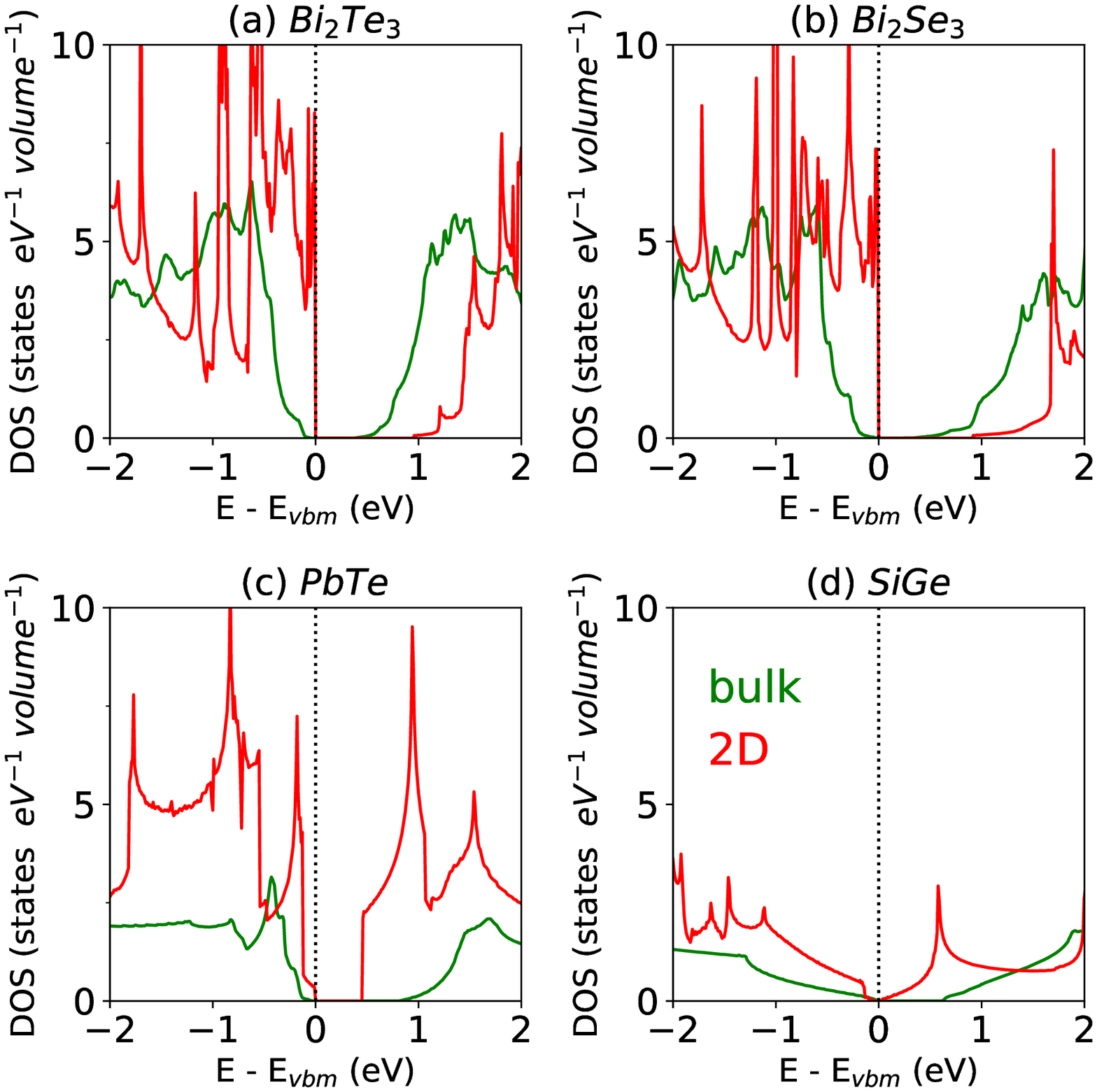}
    \caption{Density of States (DOS) of (a) $\bite$, (b) $\bise$, (c) SiGe, and (d) PbTe. The
    green colour represents the DOS of the bulk structures and the red represents the 2D
    structures.}
    \label{fig.dos}
\end{figure}

\subsection{Electronic Structure}
The calculated electronic band structures of each material are shown in Fig.~\ref{fig.band}. All bulk materials except SiGe have direct band gaps, while for the 2D counterparts, only PbTe and SiGe have the direct band gaps. The siligene exhibits a Dirac cone-shaped band structure at the K point, which was previously found in \cite{Jamdagni2015} and \cite{Sannyal2019}. $\bite$
and $\bise$ possess a similar band structure due to their similarity in structure. The band
structure of both materials in bulk has a direct band gap at $\Gamma$-point, while the inclusion
of spin-orbit coupling (SOC) causes a band inversion \cite{Witting2019}. As for the single QL,
the band structure of $\bite$ without the inclusion of SOC is similar to the previous theoretical
calculation \cite{Zhou2015}, where SOC was included. 

\begin{table}[t]
    \centering
    \caption{The calculated band gap of each material}
    \scalebox{0.7}{%
    \begin{tabular}{c c c}
    \toprule
         & This Work (eV) & Previous Work (eV) \\
         & (Without SOC) & (With SOC except SiGes)\\
    \midrule
    3D Material \\
    \hdashline
        $\bite$ & 0.33 & 0.11 \cite{Ryu2016}\\
        $\bise$ & 0.28 & 0.26 \cite{Park2016}\\
        PbTe & 0.79 & 0.19 \cite{Wang2007}\\
        SiGe & 0.6 & 1.018 \cite{Zhao2016}\\
     \midrule
    2D Material \\
    \hdashline
        $\bite$ & 0.94 & 0.32 \cite{Zhou2015}\\
        $\bise$ & 0.91 & -\\
        PbTe  & 0.45 & 0.11 \cite{Jia2017}\\
        SiGe  & 0.0052 & 0.012 \cite{Sannyal2019}\\
    \bottomrule
    \end{tabular}}
    \label{tab:gap}
\end{table}

The band gap values of each material are presented in Table
~\ref{tab:gap}. Comparing with the references, we can see that the inclusion of SOC
reduces the conduction band energy, especially in materials consisting
of heavy atoms. The band energy reduction results in the lowering of the band gap 
and band inversion in some cases, like $\bite$. SOC does not affect SiGe
tremendously because $\sige$ is composed of light atoms. 
We note that the GGA underestimates the semiconductor band gaps, 
which raises a discrepancy between this work and the reference that uses 
the GGA+U method. The total density of states (DOS) for all materials are shown
in Fig. ~\ref{fig.dos}. The energy is shifted to the valence band maximum to set it
as the reference. In all cases, the DOS near the valence
band edge is larger and denser for the 2D structures than the bulk.

\subsection{Thermoelectric Properties} 

\begin{figure*}[t]
    \centering
    \includegraphics[scale=0.5]{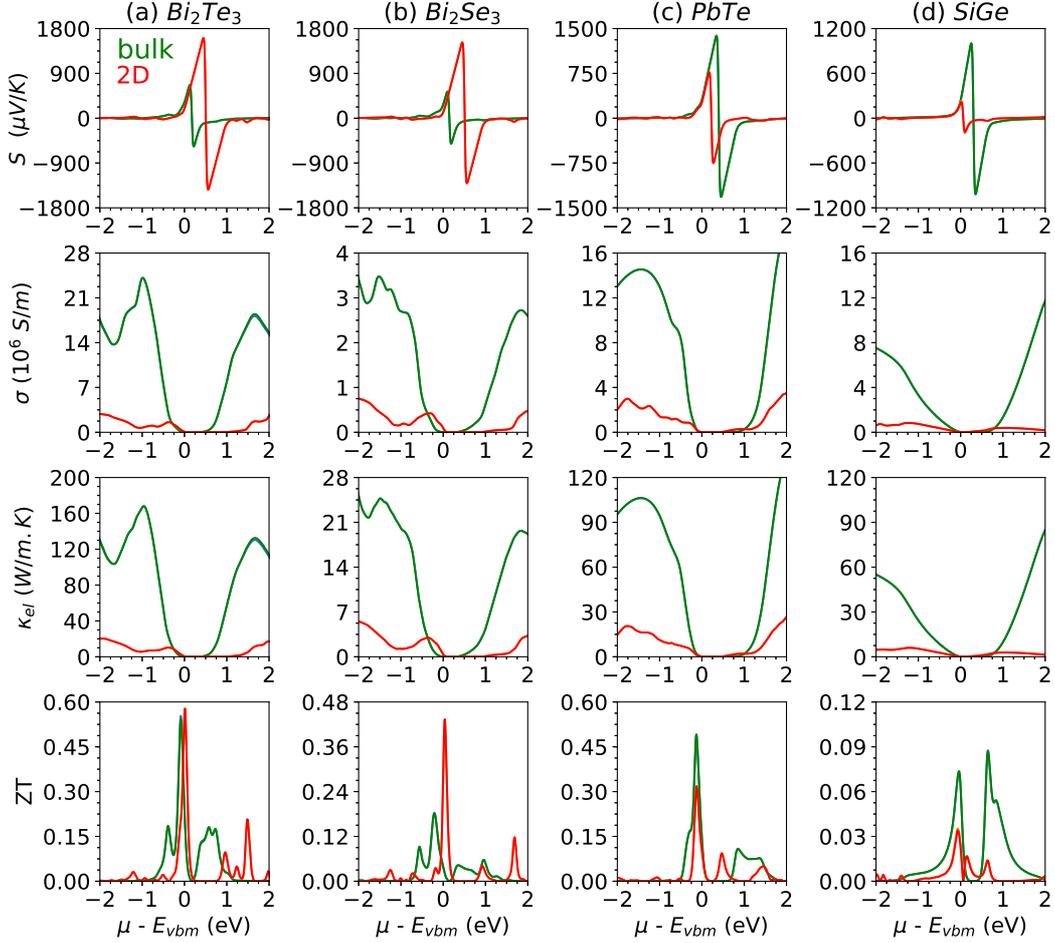}
    \caption{Thermoelectric Properties for (a) $\bite$, (b) $\bise$, (c) PbTe, (d) SiGe. The
    properties from the first row to the last: Seebeck Coefficient, Electric Conductivity,
    Electronic Thermal Conductivity, and $ZT$ Values. Bulk and 2D properties are indicated by the
    green line and red line respectively.}
    \label{fig.te}
\end{figure*}

The calculated Seebeck coefficients as a function of chemical
potential at 300 K are shown in Fig.~\ref{fig.te}. This study focuses only
on longitudinal transports to compare the bulk with the
two-dimensional properties. The properties of all materials are
isotropic. The chemical potential is related to the carrier
concentration. Increasing the chemical potential or
carrier concentration way above the gap will decrease the Seebeck coefficient.

The single QL of $\bite$ and $\bise$ have a higher Seebeck coefficient
compared to the bulk properties. On the contrary, the bulk properties
of SiGe and $\pbte$ have a much higher Seebeck coefficient than the 2D
counterparts. Single QL of $\bite$ achieved the highest Seebeck coefficient for the 
2D materials and PbTe exhibits the highest Seebeck coefficient for the 3D materials, 
with a value
of $1610\ \rm \mu V/K$ and $1375\ \rm \mu V/K$ respectively. 

The second row of Fig.~\ref{fig.te} shows the calculated electrical
conductivity as a function of chemical potential at 300K. Unlike the
Seebeck coefficient, increasing the chemical potential results in the
increase of electrical conductivity. P-type $\bite$ has the highest
bulk electrical conductivity ($\sim 24\times 10^6\ \rm S/m$) while
n-type PbTe has the highest conductivity among the 2D materials ($\sim 3\times
10^6\ \rm S/m$). $\bise$ has the smallest magnitude of electrical
conductivity. Overall, the electrical conductivities of the 2D
structures are lower compared to their bulk structures.

The calculated electronic thermal conductivities are shown in the third
row of Fig.~\ref{fig.te}. Comparing with electrical conductivity,
the thermal conductivity of each material has  similar trends. From the
second and third row, we can see that the increase of electrical
conductivity also increases electronic thermal
conductivity, which aligns with the Wiedemann-Franz law.

The $ZT$ values are shown in the last row of Fig.~\ref{fig.te}. We can
see that SiGe has the lowest maximum $ZT$ values, which are around 0.09
for the n-type bulk SiGe and 0.04 for the p-type 2D SiGe. The low $ZT$ 
value in bulk SiGe is due
to the high phonon thermal conductivity that it exhibits. The single
QL of $\bise$ and $\bite$ achieves a higher maximum $ZT$ value than the
bulk structure. The highest $ZT$ value is achieved by p-type $\bite$ on
its bulk ($\sim0.54$) and 2D structure ($\sim0.57$). 
The 2D materials do not necessarily improve the $ZT$ value. $\bise$ and $\bite$ get the $ZT$
enhancement due to the enhancement in their Seebeck coefficients,
while there are materials with lower $ZT$ values than its bulk
structure such as SiGe and PbTe.

\begin{figure}[t] 
    \centering
    \includegraphics[scale=0.38]{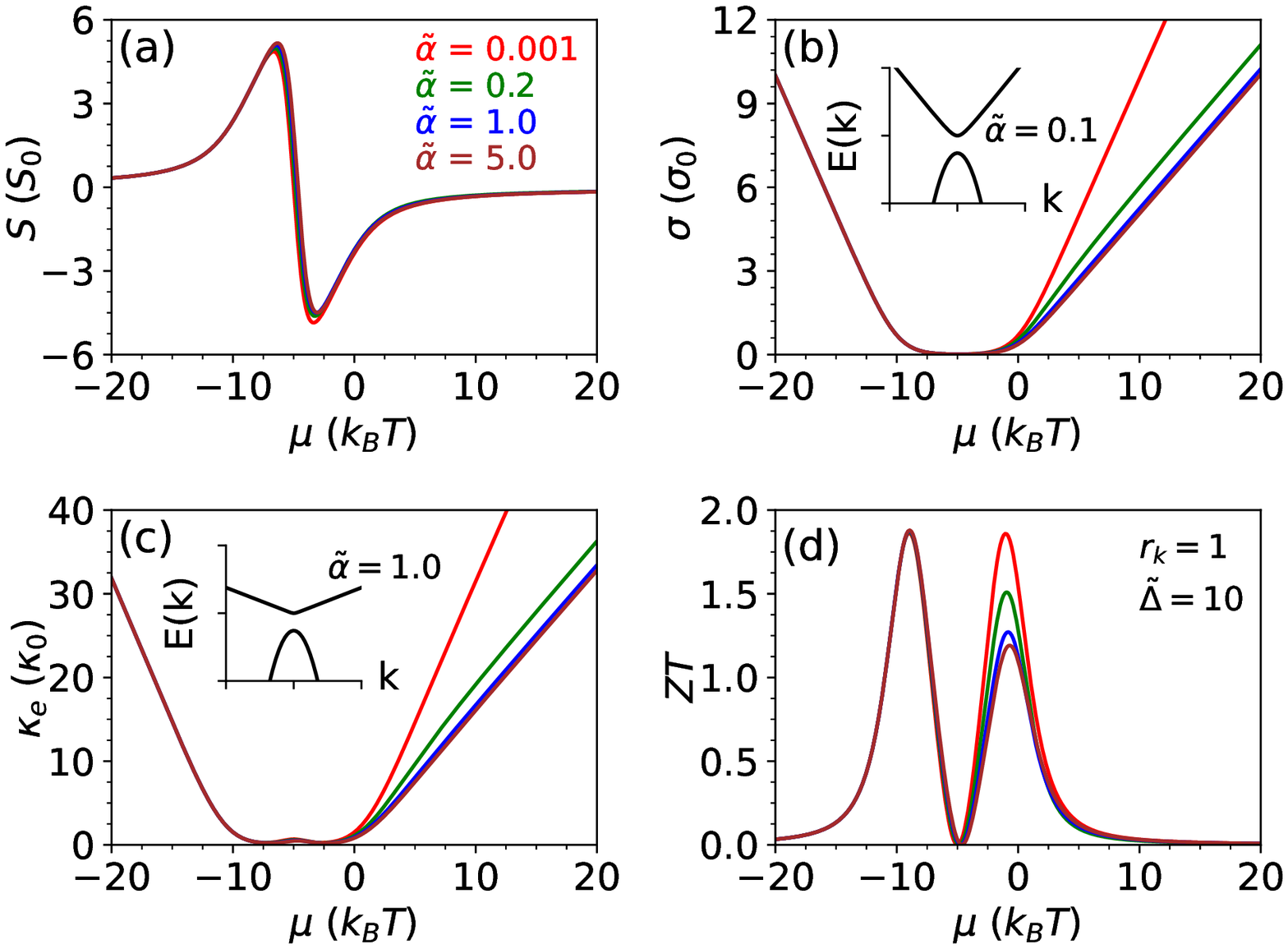}
    \caption{(a) Seebeck coefficient, (b) electrical conductivity, (c) electronic thermal conductivity, and (d) $ZT$ calculated from the 2D two-band model with $\tilde{\Delta} = 10\ k_B T$ and several band flatness. The phonon thermal conductivity $\kappa_{ph} = \kappa_0 \ r_k$ is set to be $\kappa_0$. The units are defined as follows, $S_0 = \frac{k_B}{q}$, $\sigma_0 = \frac{Cq^2}{2\pi \hbar^2}(k_B T)$, and $\kappa_0 = \frac{C k^3_B T^2}{4\pi \hbar^2}$. The inset gives the band dispersions on $\tilde{\alpha}=0.1$ and $\tilde{\alpha}=1.0$}
    \label{fig.transportprop}
\end{figure}

The band flatness and the band gap are changed upon dimensional reduction, as seen in Fig.~\ref{fig.band}. To
investigate the effects they have on transport properties, we calculate the transport
properties using a two-band model, with a Kane band as the conduction band and a parabolic band as
the valence band to emulate the asymmetrical bands near the Fermi level. The formulation can be
seen in Appendix B of Supplementary Material, and the results are shown in Fig.~\ref{fig.transportprop}, Fig.~\ref{fig.kane} and Fig. 10 - 12 in Supplementary Material.

From this simple model, we can see that band flatness gives a positive
enhancement to the TE properties in 3D material while
it has a detrimental effect on 2D material.  Changing the band flatness will only affect the Kane band in CB, thus we only plot $ZT_{max}$ vs band gap $\tilde{\Delta}$ for n-doped only (Fig. \ref{fig.kane}). Band flatness does not affect the Seebeck coefficient significantly, but rather it affects $\sigma$ and $\kappa_{el}$ more (Fig. 10 - 12). In 2D systems, $\sigma$ and $\kappa_{el}$ decrease as the band becomes flatter while the Seebeck coefficient remains the same. As a result, the $ZT$ in 2D materials possessing a flat band is lower than those with a more dispersive band. On the contrary, band flatness has the opposite effect on TE properties in 3D due to different DOS. Aside from band flatness, the asymmetrical effective mass parameter described in Ref.~\cite{Markov2019} might affect TE properties. However, in this work, we assume the masses to be the same. This ratio only affects the 3D systems and has no effects on the 2D, because there is no mass terms in the 2D TE integral (Eq. 18-21 of Supplemental Material).

In general, the maximum $ZT$ value ($ZT_{max}$) increases proportionally with the band gap in both the 3D and 2D materials. The maximum $ZT$ values increase as the band gap widens up to a
certain threshold value, which is around $10 k_{B} T$ or 0.25 eV at room temperature, and become
saturated beyond this value. The optimum band gap that we obtain is the same as the previous
works \cite{Hasdeo2019, Mahan1994}. 

From our two band model, we can explain the first-principles calculation results. The 2D PbTe has a low $ZT$ value because it possesses flat bands near its band
edge, while 2D SiGe has a low band gap resulting in a low $ZT$. On the other hand, $\bite$ and $\bise$ bands are corrugated near the Fermi level and are
not classified as the flat band as described by the Kane model, so the results from our model cannot be
used to describe these materials. The Corrugated flat band has multiple Fermi pockets that, in
effect, enhance the Seebeck coefficients \cite{Mori2016}. We also note that in Fig. ~\ref{fig.dos},  there are a lot of sharp peaks in 2D $\bite$ and 2D $\bise$ DOS, while 2D SiGe and 2D PbTe have less of them. 

Our continuum model is not able to explain the effect of narrow band width on TE properties. According to the previous works \cite{Mahan1996}, the upper limit of $ZT$ is achieved by having a transport distribution function (TDF) that resembles the Dirac delta function. However, according to \cite{Chen2011, Jeong2012}, such TDF cannot be achieved in the real material. Even if the DOS shows the Van Hove singularities, the TDF is not divergent because the DOS term is canceled out with the square of longitudinal velocity term resulting in a finite $ZT$ \cite{Nurhuda_2020}. The narrow transport distribution gives more conducting channels that increase $\sigma$ while giving a low $\kappa_{el}$ because of the $(E-\mu)^2$ factor that $\kappa_{el}$ has \cite{Chen2011, Jeong2012}. Sharp peaks in DOS have also been found in previous works \cite{Xue2015, Ding2019}, which enhance the Seebeck coefficient. These works explain why 2D $\bite$ and 2D $\bise$, which have corrugated bands plus sharp peaks in DOS, possess a high $ZT$.

\begin{figure}[t] 
    \centering
    \includegraphics[scale=0.5]{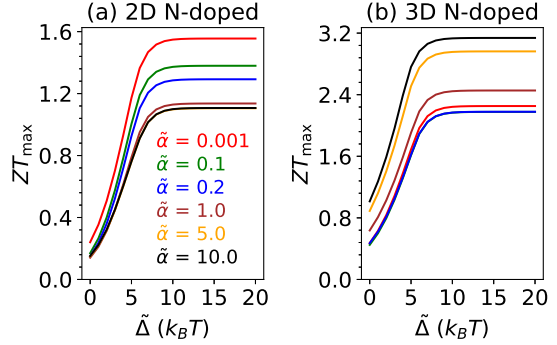}
    \caption{$ZT$ at optimal $\mu$ ($ZT_{max}$) vs band gap $\tilde{\Delta}$ for various band flatness. $ZT_{max}$ increases proportionally with $\tilde{\Delta}$ up to around 10 $k_B T$.}
    \label{fig.kane}
\end{figure}

\begin{figure*}[t] 
    \centering
    \includegraphics[scale=0.38]{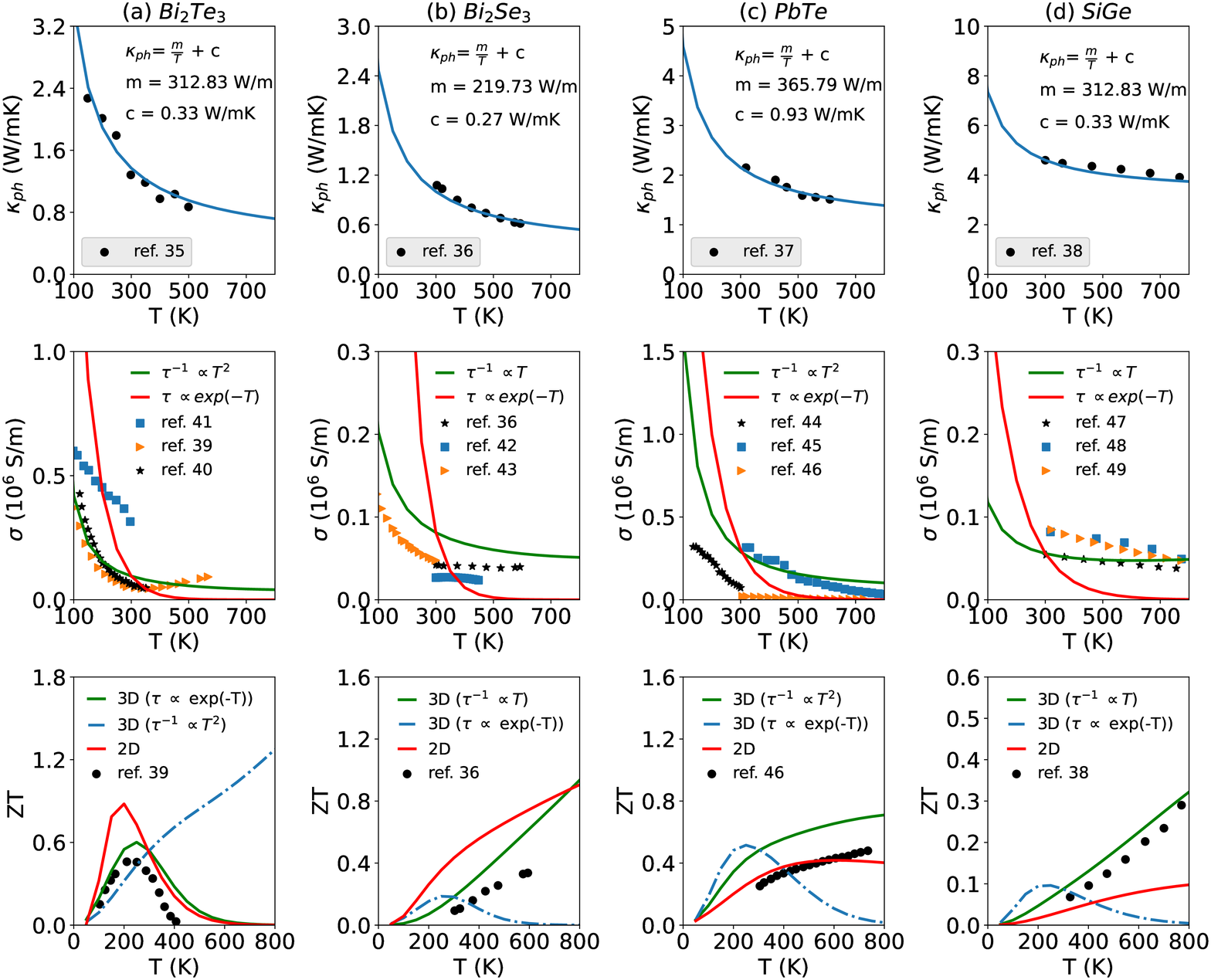}
    \caption{Phonon thermal conductivity (first row) curve fitting for (a) $\bite$ \cite{Qiu2010}, (b) $\bise$ \cite{Liu2017}, (c) PbTe \cite{Masaki2000}, and (d) SiGe \cite{Tayebi2012} as a function of temperature. The second row shows the results of the fitting from Fig. 9 in Appendix and direct fitting on electric conductivity. The temperature dependence of $ZT$ values are shown in the third row, the green line indicates that the fittest relaxation time is used, while the red line indicates the 2D $ZT$ value. All thermoelectric properties are calculated on its optimized chemical potential. }
    \label{fig.fitting}
\end{figure*}

The optimized chemical potential and its associated carrier
concentration are given in Table. 4. 
After obtaining the optimized chemical potentials, we calculate the temperature-dependent relaxation time
and phonon thermal conductivity on those chemical potentials. The
relaxation times of each material on various temperatures are obtained
using the same method to get the value in Table~\ref{tab.tau}. As for the
phonon conductivities, we obtain them directly from the experimental
data fitting. All of the phonon conductivities exhibit a $1/T$
dependency. We then try to see if the fitted relaxation time can
display a similar trend with the experimental data by comparing the
electrical conductivities (Fig.~\ref{fig.fitting} second row). It was shown that the
temperature dependency we got from Fig. 9 represents the
experimental data poorly.  Therefore, we  fit the relaxation time from
the electrical conductivities directly by using the $\sigma/\tau$ value from the calculation at the optimized chemical potential, 
which gives better results. The discrepancies occur because experimental carrier concentrations were unknown. We then compare both of relaxation time function in
the $ZT$ value (Fig.~\ref{fig.fitting} third row).

From the plot, we can see that the exponential relaxation time capture the temperature dependency
of $\bite$ data Ref.~\cite{Jeon1991}. In $\bise$ and SiGe, the relaxation time is proportional to
$1/T$, while in PbTe, it is proportional to $1/T^2$. The same temperature dependencies are used for
each corresponding 2D material due to the limited experimental data. The fittest relaxation time
functions are as follow,
\begin{align*}
        \tau_{\bite} &= 6.88\ \text{exp}(-\text{T}/54.5) \ \ \text{ps},\\
        \tau_{\bise} &= 2.098\ \text{T}^{-1} \ \ \text{ps},\\
        \tau_{\pbte} &= 0.948\ \text{T}^{-2} \ \ \text{ns},\\
        \tau_{\sige} &= 2.4\ \text{T}^{-1} \ \ \text{ps}.    
\end{align*}
In room temperature, all of the relaxation times are in the order of fs.

The $ZT$ value of a single QL $\bite$ is bigger on low temperature ($<$
300K), and the value drops beyond the room temperature like its bulk
counterpart. The single QL $\bise$ showcase a better $ZT$ value on
all temperature ranges except on $>$ 800K. For PbTe and SiGe, we can
see that the bulk $ZT$ values are higher than 2D values on
all temperature range. None of the materials reaches a $ZT$ value much
higher than unity, the highest 2D $ZT$ value is achieved by $\bite$ in
low temperature regime (around 0.88 at 200K) and $\bise$ in high
temperature regime (around 0.9 at 800K). The highest bulk $ZT$ value is
achieved by $\bise$ on 800K (around 0.93).

In conclusion, we have calculated the thermoelectric properties of the
bulk and the 2D structures of $\bite$, $\bise$, PbTe, and SiGe. We used
temperature-dependent relaxation time approximation 
to obtain the transport properties. The single QL of $\bite$ and
$\bise$ exhibits a higher $ZT$ than its bulk due to their corrugated flat band, 
which agree with HD theory. However, PbTe and
SiGe violate the HD theory. From the two-band model analysis, 2D PbTe and SiGe have lower $ZT$ than 3D counterparts because 2D PbTe has flat bands and 2D SiGe has a low
band gap. We note that  these low $ZT$ occur even when the DOS of 2D materials are higher than 3D. The fact that HD theory is non-universal requires a deeper
analysis of which material or geometry performs the best at a given dimension. 

The figure of
merits on all materials, except $\bite$, increase as the temperature
increases. The single QL of $\bite$ has a higher $ZT$ value below room
temperature, while the single QL of $\bise$ has a higher $ZT$ value on
temperature range below 800K. PbTe and SiGe monolayers have lower $ZT$ values on
all temperature ranges than their bulk. Better results might be achieved
when one can manipulate the relaxation mechanisms to reduce phonon
thermal conductivity and to increase electrical conductivity for the
bulk and the 2D structures. \textcolor{white}{\cite{Plechacek2004, Kulbachinskii2012, Hong2015, Hor2009, McGuire2008, Basu2013, Pei2012, Nozariasbmarz2016, Wang2008, Bathula2012, Tayebi2012, Janicek2009, Scheidemantel2003, Goldsmid_1958, Hunter2007, Nakajima1963, Huang2008, Dalven1969, Sharma2016, Steele1958}}

\section*{Acknowledgments}
The computation in this work has been done using the facilities of HPC LIPI, Indonesian Institute of Sciences (LIPI). EHH acknowledges ATTRACT 7556175 and CORE 11352881

\section*{References}

\providecommand{\newblock}{}

\onecolumn

\section*{Supplementary Material}

\section*{Appendix A: Relaxation Time Fitting}

Here we present the fitting method to obtain the relaxation time at various temperatures. The method is adopted from Scheidemantel's and Goldsmid's work \cite{Scheidemantel2003, Goldsmid_1958}. All the fitting results are shown in Fig. 8. We plot the Seebeck coefficient with respect to the electrical conductivity on each chemical potential at a certain temperature from BoltzTraP outputs. The electrical conductivity from BoltzTraP is in the form of $S/\tau$, thus we can compare the plot with experimental data to obtain the relaxation time on a certain temperature. 

From the fitting results in Fig. 8, we can do an additional curve fitting to obtain the temperature dependency (Fig. 9). It is shown that all of the material has an exponential dependency on the temperature. But it is shown in Fig. 7, that they describe the conductivity poorly. So we tried to do a direct curve fitting with the conductivity to obtain a better relaxation time. Experiment data from \cite{Plechacek2004, Hor2009, Basu2013, Nozariasbmarz2016} are used for the direct curve fitting of $\bite$, $\bise$, PbTe,  and SiGe, respectively. We fixed the relaxation time on 300 K to the values in Table.~\ref{tab.tau} on the direct curve fitting to attain the same thermoelectric properties we have calculated on Fig.~\ref{fig.te}.

\section*{Appendix B: Asymmetric Bands}

Here we present the formulation we used to calculate the thermoelectric properties of asymmetric bands. We used two-bands model, with parabolic band as the valence band and Kane band as the conduction band, both are defined as,
\begin{align}
    E^{cb}_{Kane} &= \sqrt{\frac{\hbar^2 k^2}{2m\alpha} + \frac{1}{4\alpha^2}} - \frac{1}{2\alpha},\\
    E^{vb}_{parabolic} &= - \frac{\hbar^2 k^2}{2m} - \Delta,
\end{align}
where $\Delta$ is the band gap and $\alpha$ is the non-parabolicity factor. The value of $\alpha = 0$ corresponds to a parabolic band.
The transport properties from Boltzmann's transport theory under relaxation time approximation (RTA) , for each band, are given by,

\begin{align}
    \sigma &= q^2 \mathcal{L}_0,\\
    S &= \frac{1}{qT} \frac{\mathcal{L}_1}{\mathcal{L}_0},\\
    \kappa_e &= \frac{1}{T} \Big( \mathcal{L}_2 - \frac{(\mathcal{L}_1)^2}{\mathcal{L}_0}\Big),\\
    ZT &= \frac{\sigma S^2}{\kappa_e + \kappa_{ph}} T,
\end{align}

where $\mathcal{L}_i$ is the TE integral and is defined as,

\begin{align}
    \mathcal{L}_{i,vb} = \int_{-\infty}^{0} \tau(E) (E-\mu)^i \Big(-\frac{\partial{f}}{\partial{E}}\Big)dE,\\
    \mathcal{L}_{i,cb} = \int_{0}^{\infty} \tau(E) (E-\mu)^i \Big(-\frac{\partial{f}}{\partial{E}}\Big)dE,
\end{align}

where $\mu$ is the Fermi energy, $f(E)$ is the Fermi-Dirac distribution, and $\tau(E)$ is the transport distribution function (TDF). The explicit form of TDF with constant relaxation time approximation (CRTA) is $\tau (E) = v^2_x(E) \origtau (E) D(E)$ and has a different form on each band dispersion and dimension. The TDF that are used in the calculations are:

\begin{align}
    \tau (E)_{\text{kane}}^{2D} &= C \Big(\frac{1}{4m \alpha}\Big) \frac{E(E+\frac{1}{\alpha})} {(E + \frac{1}{2 \alpha})^2} D(E)_{\text{kane}}^{2D},\\
    \tau (E)_{\text{kane}}^{3D} &= C \Big(\frac{1}{6m \alpha}\Big) \frac{E(E+\frac{1}{\alpha})} {(E + \frac{1}{2 \alpha})^2} D(E)_{\text{kane}}^{3D},\\
    \tau (E)_{\text{parabolic}}^{2D} &= C \Big(\frac{-E-\Delta}{m}\Big) D_{\text{parabolic}}^{2D},\\
    \tau (E)_{\text{parabolic}}^{3D} &= C \Big(\frac{-E-\Delta}{m}\Big) D_{\text{parabolic}}^{3D}.
\end{align}

The DOS of each band dispersions and dimensions can be written as,

\begin{align}
    D(E)_{\text{kane}}^{2D} &= \frac{m \alpha}{\pi \hbar^2} \Big(E+\frac{1}{2 \alpha} \Big),\\
    D(E)_{\text{kane}}^{3D} &= \frac{1}{2 \pi^2} \Big(\frac{2 m \alpha}{\hbar^2}\Big)^{3/2} \Big(E+\frac{1}{2\alpha}\Big) \Big(E \Big(E+\frac{1}{\alpha} \Big)\Big)^{1/2},\\
    D(E)_{\text{parabolic}}^{2D} &= \frac{m}{2\pi\hbar^2}\Theta(|E| - |\Delta|),\\
    D(E)_{\text{parabolic}}^{3D} &= \frac{\sqrt2 m^{3/2}}{2\pi^2\hbar^3} (-E-\Delta)^{1/2} \ \Theta(|E|-|\Delta|),
\end{align}

where $\Theta$ is the Heaviside function. Defining the dimensionless quantity as
\begin{align*}
    E &= \epsilon k_B T,\\
    \mu &= \eta k_B ,\\
    \Delta &= \tilde{\Delta} k_B T,\\
    \alpha &= \frac{\tilde{\alpha}}{k_B T},
\end{align*}
plus letting $x = \epsilon - \eta$, the TE integral then become:

\begin{align}    
    \mathcal{L}_{i,c}^{2D} &= \frac{C}{4\pi \hbar^2}(k_B T)^{i+1} \mathcal{H}_{i,c}(\eta),\\
    \mathcal{L}_{i,v}^{2D} &= \frac{C}{2\pi \hbar^2}(k_B T)^{i+1} \Big[ \mathcal{F}_{i+1,v}(\eta+\tilde{\Delta}) + (\eta + \tilde{\Delta} \mathcal{F}_{i,v}(\eta + \tilde{\Delta}) \Big],\\
    \mathcal{L}_{i,c}^{3D} &= \frac{C(2m)^{1/2}}{6\pi^2 \hbar^3}(k_B T)^{i+3/2} \ \mathcal{I}_{i,c}(\eta),\\
    \mathcal{L}_{i,v}^{3D} &= \frac{C(2m)^{1/2}}{3\pi^2 \hbar^3}(k_B T)^{i+3/2} \ \mathcal{J}_{i,v}(\eta + \tilde{\Delta}),
\end{align}

where $\mathcal{H}_{i,c}(\eta)$, $\mathcal{F}_{i,v}(\eta)$, $\mathcal{I}_{i,c}(\eta)$, $\mathcal{J}_{i,v}(\eta)$ are defined as:

\begin{align}
    \mathcal{H}_{i,c}(\eta) &= \int_{-\eta}^{\infty} x^i \frac{(x+\eta) (x+\eta+1/\tilde{\alpha})} {(x+\eta+1/\tilde{2 \alpha})} \frac{exp(x)}{(exp(x)+1)^2} dx,\\
     \mathcal{F}_{i,v}(\eta) &= - \int_{-\infty}^{-\eta} x^i  \frac{exp(x)}{(exp(x)+1)^2} dx,\\
    \mathcal{I}_{i,c}(\eta) &= \tilde{\alpha}^{1/2} \int_{-\eta}^{\infty} x^i \Big[ \frac{(x+\eta)^3 (x+\eta+1/\tilde{\alpha})^3} {(x+\eta+1/\tilde{2 \alpha})^2} \Big]^{1/2} \frac{exp(x)}{(exp(x)+1)^2} dx,\\
    \mathcal{J}_{i,v}(\eta) &= \int_{-\infty}^{-\eta} x^i (-x-\eta-\tilde{\Delta})^{3/2} \frac{exp(x)}{(exp(x)+1)^2} dx.
\end{align}

Only $\mathcal{F}_{i,v}(\eta)$ can be solved analytically out of the four integrals. The analytic results for these integrals are:

\begin{align}
    \mathcal{F}_{0,v}(\eta) &= \frac{1}{e^{\eta}+1},\\
    \mathcal{F}_{1,v}(\eta) &= -\frac{\eta}{e^{\eta}+1} - ln(1+e^{-\eta}),\\
    \mathcal{F}_{2,v}(\eta) &= \frac{\eta^2}{e^{\eta}+1} + 2 \eta ln(1+e^{-\eta}) - 2Li_2(-e^{-\eta}),\\
    \mathcal{F}_{3,v} (\eta) &= \eta^2\left(\frac{\eta}{1+e^\eta}+3\ln\left(1+e^{-\eta}\right)\right)-6\eta\mathrm{Li}_2(-e^{-\eta})-6\mathrm{Li}_3(-e^{-\eta}) 
\end{align}

with $Li_k(z) = \sum_{n=1}^{\infty} \frac{z^n}{n^k}$.\\

We can obtain thermoelectric properties by plugging eq. (18) - (21) to eq. (4) - (6). The transports from conduction band have the form of:

\begin{align}
    \sigma^{2d}_c &= \frac{Cq^2}{4\pi\hbar^2}(k_B T) \mathcal{H}_{0,c}(\eta) = \sigma^{2d}_c = \sigma^0_c \tilde{\sigma}^{2d}_c \implies \sigma^0_c = \frac{Cq^2}{4\pi\hbar^2}(k_B T),\\
    S^{2d}_c &= - \frac{k_B}{q} \frac{\mathcal{H}_{1,c}(\eta)}{\mathcal{H}_{0, c}(\eta)} = - S^0_c \tilde{S}^0_c \implies S^0_c = \frac{k_B}{q},\\
    \kappa^{2d}_{e,c} &= \frac{C k^3_B T^2}{4 \pi \hbar^2} \Big( \mathcal{H}_{2,c} - \frac{(\mathcal{H}_{1,c})^2}{\mathcal{H}_{0,c}} \Big) = \kappa^0_{e,c} \tilde{\kappa}^0_{e,c} \implies \kappa^0_{e,c} = \frac{C k^3_B T^2}{4 \pi \hbar^2},
\end{align}

while the transports from the valence have the form of:

\begin{align}
\begin{split}
    \sigma^{2d}_v &= \frac{Cq^2}{2\pi\hbar^2}(k_B T) \Big[ \mathcal{F}_{1,v}(\eta + \tilde{\Delta}) + (\eta + \tilde{\Delta}) \mathcal{F}_{0,v} (\eta + \tilde{\Delta}) \Big]\\
    &= \sigma^0_v \tilde{\sigma}^{2d}_v \implies \sigma^0_v = \frac{Cq^2}{2\pi\hbar^2}(k_B T),\\
\end{split}
\end{align}

\begin{flalign}
\begin{split}
    S^{2d}_v &= - \frac{k_B}{q} \Big[ \frac{\mathcal{F}_{2,v}(\eta + \tilde{\Delta}) + (\eta + \tilde{\Delta}) \mathcal{F}_{1,v} (\eta + \tilde{\Delta})} {\mathcal{F}_{1,v}(\eta + \tilde{\Delta} + (\eta + \tilde{\Delta}) \mathcal{F}_{0,v} (\eta + \tilde{\Delta})} \Big] \\
    &= - S^0_v \tilde{S}^0_v \implies S^0_v = \frac{k_B}{q},\\
\end{split}
\end{flalign}

\begin{align}
\begin{split}
    \kappa^{2d}_{e,v} &= \frac{C k^3_B T^2}{4 \pi \hbar^2} \Big( \mathcal{H}_{2,c} - \frac{(\mathcal{H}_{1,c})^2}{\mathcal{H}_{0,c}} \Big) \\
    &= \frac{C k^3_B T^2}{4 \pi \hbar^2}  \Bigg[ \Big( \mathcal{F}_{3,v}(\eta + \tilde{\Delta}) + (\eta + \tilde{\Delta}) \mathcal{F}_{2,v} (\eta + \tilde{\Delta}) \Big) \\
    & \qquad \qquad - \Big( \frac{\mathcal{F}_{2,v}(\eta + \tilde{\Delta}) + (\eta + \tilde{\Delta}) \mathcal{F}_{1,v} (\eta + \tilde{\Delta})}{\mathcal{F}_{1,v}(\eta + \tilde{\Delta}) + (\eta + \tilde{\Delta}) \mathcal{F}_{0,v} (\eta + \tilde{\Delta})} \Big) \Bigg]\\
    &= \kappa^0_{e,v} \tilde{\kappa}^0_{e,v} \implies \kappa^0_{e,v} = \frac{C k^3_B T^2}{4 \pi \hbar^2}.
    \end{split}
\end{align}

Looking at the transport quantities from eq. (30) - (35), the relations between the conduction band and the valence band are

\begin{align*}
    \sigma^0_c &= \frac12 \sigma^0_v,\\
    S^0_c &= S^0_v,\\
    \kappa^0_{e,c} &= \frac12 \kappa^0_{e,v}.
\end{align*}

The 3D formulation retains the same relation as above, and the transports equations are the same as the 2D formulation, differing only in the transport magnitude:

\begin{align*}
    \sigma^0_{c,3D} &= \frac{Cq^2 (2m)^{1/2}}{6 \pi \hbar^2} (k_B T)^{3/2}\\
    S^0_{c,3D} &= \frac{k_B}{q}\\
    \kappa^0_{e,c,3D} &= \frac{C(2m)^{1/2}}{6 \pi^2 \hbar^3} k^{7/2}_B T^{5/2}
\end{align*}

For the two-band model, the total transport properties are:
\begin{align}
    \begin{split}
    \sigma &= \sigma^{c} + \sigma^{v} \\
    &= \sigma^0_v \Big(\frac12 \tilde{\sigma}^0_c + \tilde{\sigma}^0_v \Big)\\
    &= \sigma^0_v \tilde{\sigma}.
    \end{split}
\end{align}
    
\begin{align}
    \begin{split}
    S &= \frac{\sigma^{c} S^{c} + \sigma^{v} S^{v}}{\sigma^c + \sigma^v} \\
    &= S^0_v \Big[ \frac{\frac12 \tilde{\sigma}^0_c \tilde{S}^0_c + \tilde{\sigma}^0_v \tilde{S}^0_v}{\frac12 \tilde{\sigma}^0_c + \tilde{\sigma}^0_v} \Big] \\
    &= S^0_v \tilde{S}.
    \end{split}
\end{align}

\begin{align}
    \begin{split}
    \quad \qquad \qquad \qquad \qquad \kappa_e &= \frac{\sigma^c \sigma^v}{\sigma^c + \sigma^v} (S^c - S^v)^2 + (\kappa^c_e + \kappa^v_e) \\
    \quad \qquad \qquad \qquad \qquad &= \kappa^0_{e,v} \Big[ \frac{\frac12 \tilde{\sigma}^0_c \tilde{\sigma}^0_v} {\frac12 \tilde{\sigma}^0_c + \tilde{\sigma}^0_v} \big( \tilde{S}^0_c - \tilde{S}^0_v \big)^2 + \Big( \frac12 \tilde{\kappa}_{e,0}^c + \tilde{\kappa}_{e,0}^v \Big) \Big]\\
    \quad \qquad \qquad \qquad \qquad &= \kappa^0_{e,v} \tilde{\kappa_e}.
    \end{split}
\end{align}
\begin{align}
    \begin{split}
    ZT &= \frac{\sigma S^2}{\kappa_e + \kappa_{ph}} \\
    &= \frac{\tilde{\sigma} \tilde{S}^2}{\tilde{\kappa}_e + r_k}.
    \end{split}
\end{align}
Phonon thermal conductivity is defined as $\kappa_{ph} = r_k \kappa^0_{e,v}$ in the equations above. We use $r_k = 1$ in all our calculations.\\

Each transport properties from multiple bands can be written as the summation of the kernel integrals, with $n$ as the total number of bands,

\begin{align}
    \sigma &= q^2 \sum_{i=1}^n \mathcal{L}_{0, i} = \sum_{i=1}^n \sigma_i,\\
    S &= \frac{1}{qT}  \frac{ \sum_{i=1}^n \mathcal{L}_{1,i}}{  \sum_{i=1}^n\mathcal{L}_{0,i}} = \frac{ \sum_{i=1}^n S_i \sigma_i}{ \sum_{i=1}^n \sigma_i}
\end{align}

\begin{align}
    \begin{split}
    \kappa_e &= \frac{1}{T} \Big(  \sum_{i=1}^n \mathcal{L}_{2,i} - \frac{( \sum_{i=1}^n \mathcal{L}_{1,i})^2}{ \sum_{i=1}^n \mathcal{L}_{0,i}}\Big) \\
    &=  \sum_{i=1}^n \kappa_{e,i} +  \sum_{\substack{i, j \\ i \neq j}}^n \frac{\sigma_i \sigma_j}{\sigma_i + \sigma_j} (S_i - S_j)^2
    \end{split}
\end{align}

\newpage

\begin{figure}[h] 
    \centering
    \includegraphics[scale=0.4]{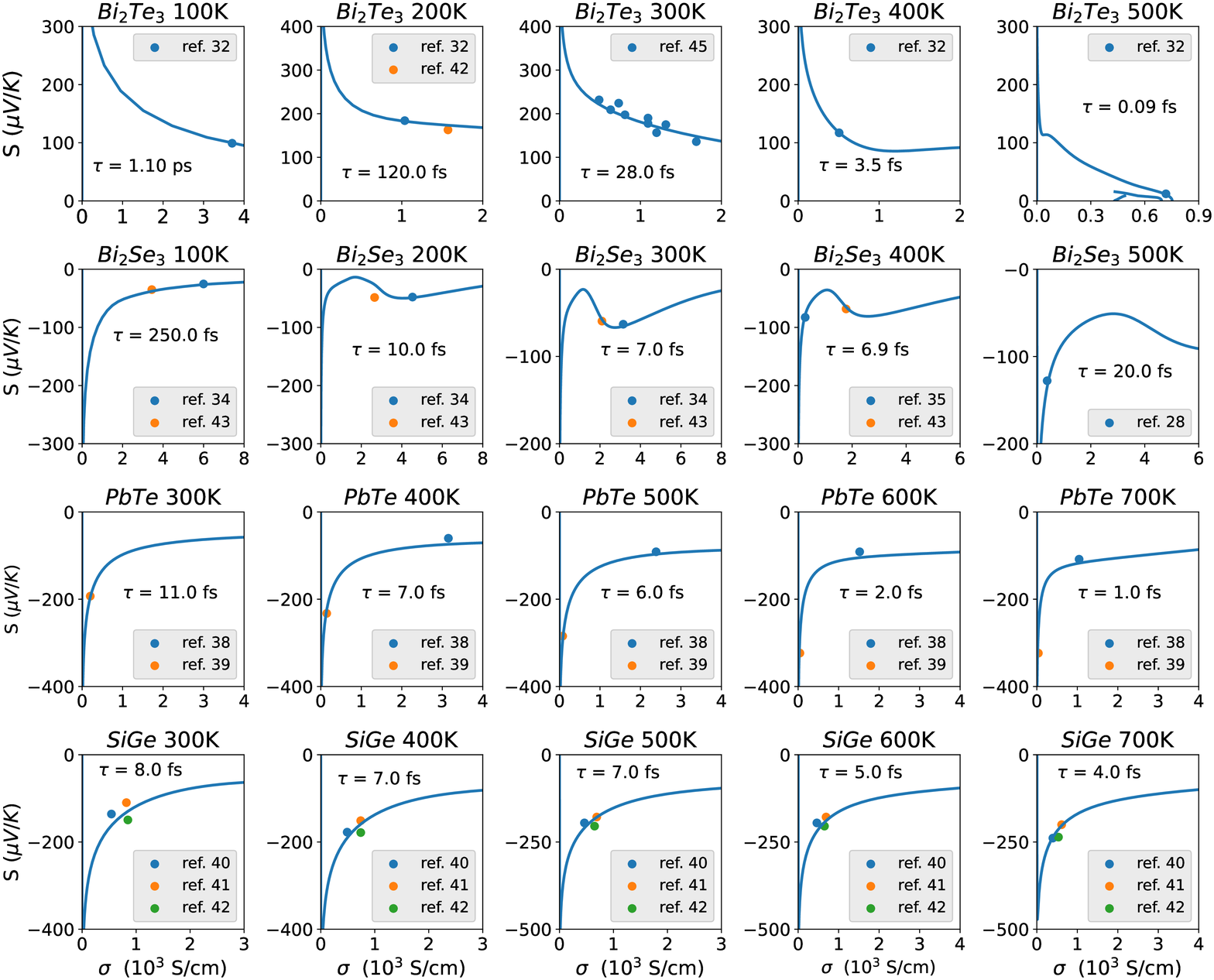}
    \caption{Comparison of Seebeck coefficient versus electrical conductivity from the BoltzTraP outputs and experimental data to obtain the relaxation time at various temperature. }
    \label{fig.relaxation_fitting}
\end{figure}

\begin{figure}[h] 
    \centering
    \includegraphics[scale=0.35]{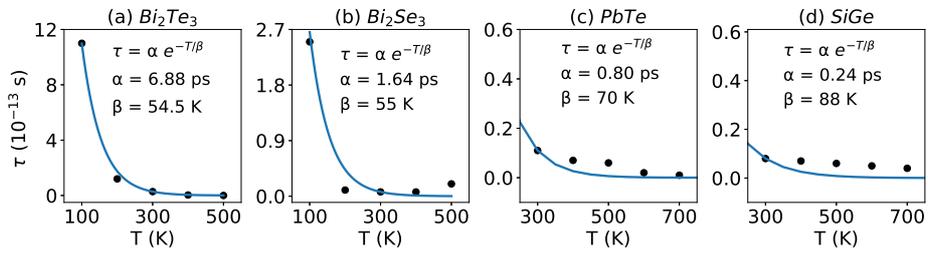}
    \caption{Temperature dependency of the relaxation time obtained from curve fitting. Each data is the result of the curve fitting in Fig.~\ref{fig.relaxation_fitting}}
    \label{fig.relax_vs_tmp}
\end{figure}

\begin{figure}[h] 
    \centering
    \includegraphics[scale=0.35]{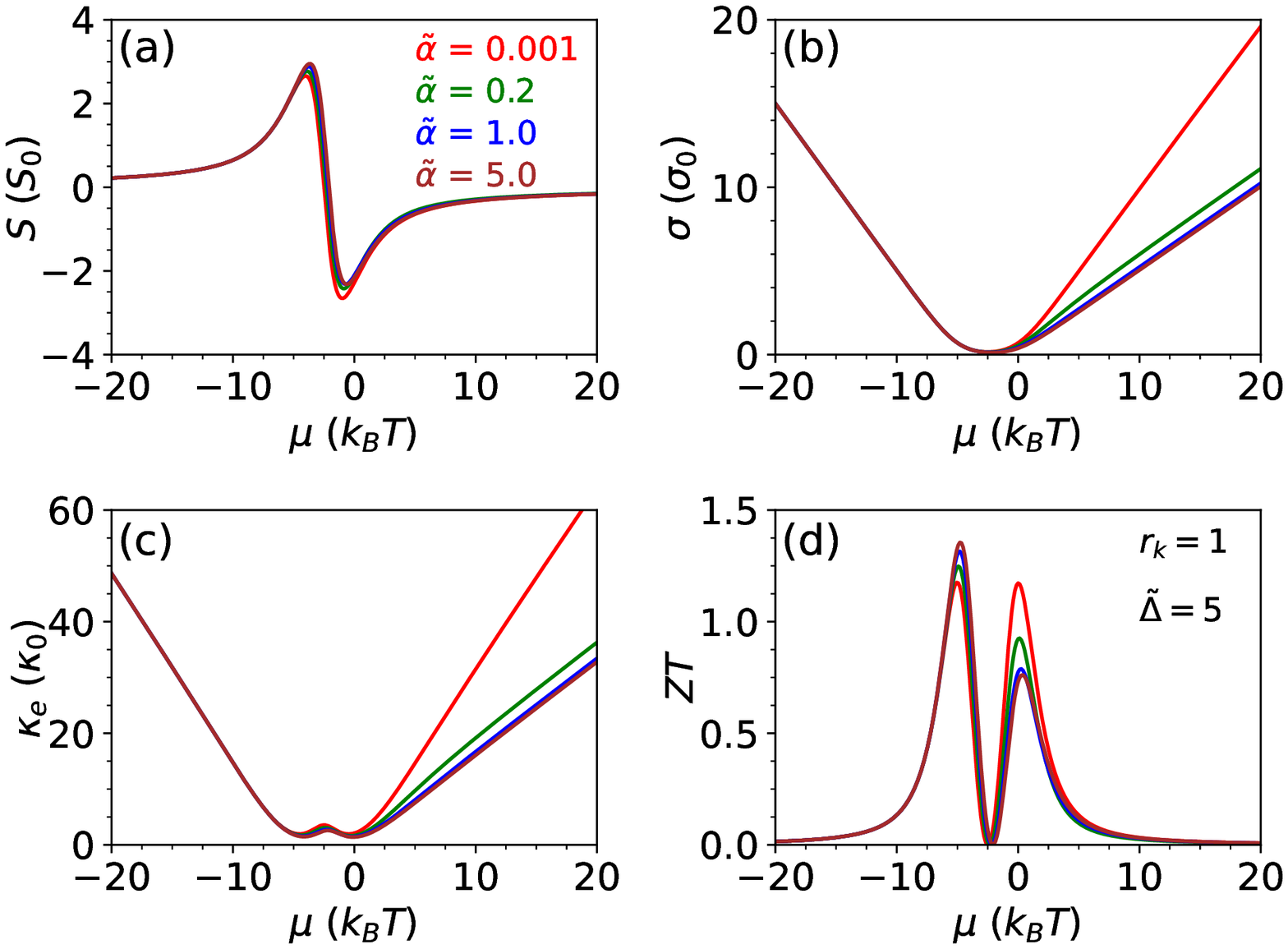}
    \caption{Thermoelectric properties of 2D two-bands model on several band flatness with $\tilde{\Delta} = 5$ and $r_k = 1$.}
    \label{fig.2d_delta5}
\end{figure}

\begin{figure}[h] 
    \centering
    \includegraphics[scale=0.35]{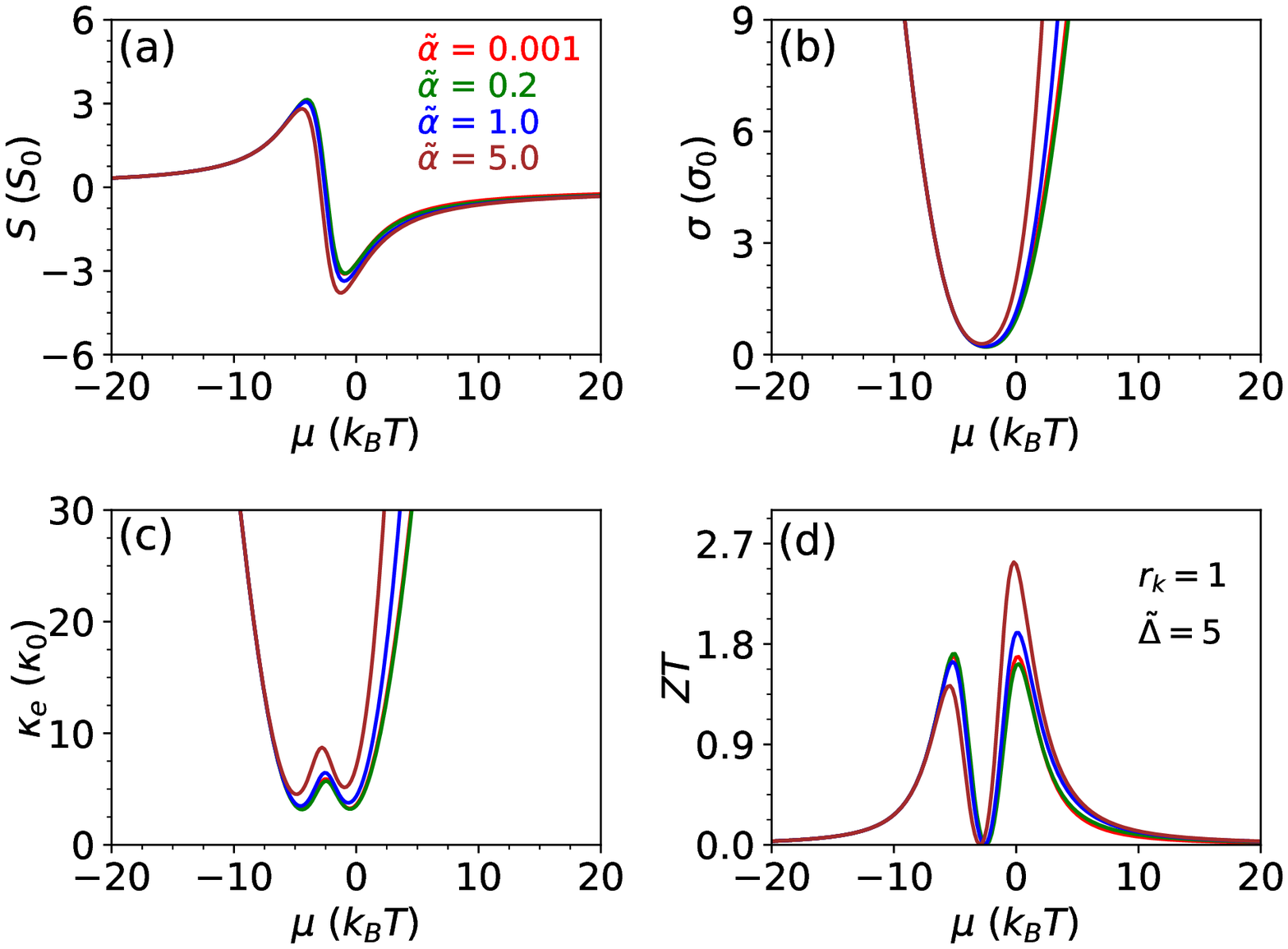}
    \caption{Thermoelectric properties of 3D two-bands model on several band flatness with $\tilde{\Delta} = 5$ and $r_k = 1$.}
    \label{fig.3d_delta5}
\end{figure}

\begin{figure}[h] 
    \centering
    \includegraphics[scale=0.35]{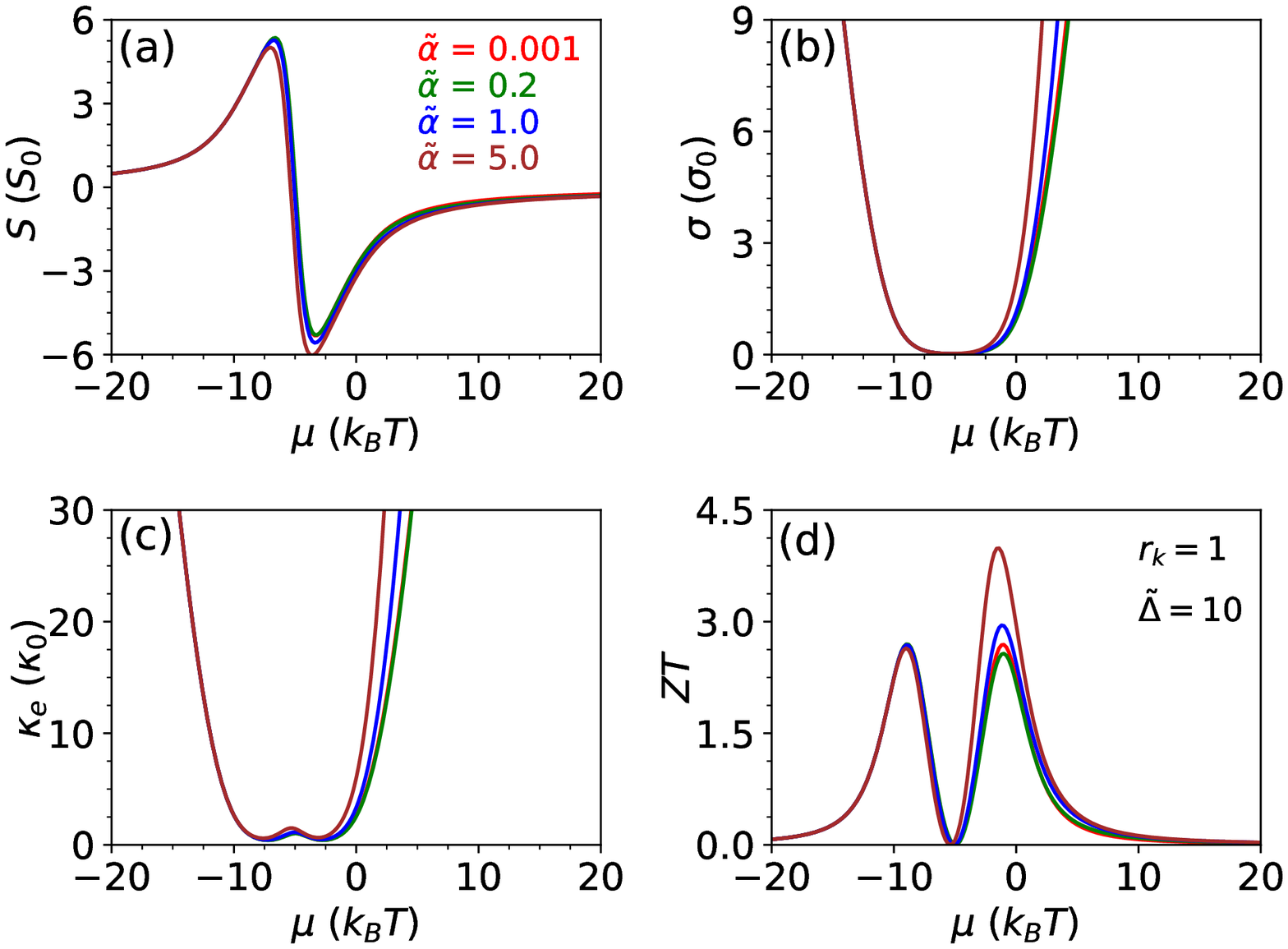}
    \caption{Thermoelectric properties of 3D two-bands model on several band flatness with $\tilde{\Delta} = 10$ and $r_k = 1$.}
    \label{fig.3d_delta10}
\end{figure}

\newpage
-
\begin{table}[h]
    \centering
    \caption{Relaxed lattice constants}
    \scalebox{0.7}{%
    \begin{tabular}{c c c c}
    \toprule
         & $a$ = $b$ ($\textrm{\r{A}}$) & Experiment ($\textrm{\r{A}}$) & Computation ($\textrm{\r{A}}$) \\
    \midrule
    3D Material \\
    \hdashline
        $\bite$ & 10.6555 & 10.476 [47] \textcolor{white}{ \cite{Nakajima1963}} & 10.473  [48] \textcolor{white}{\cite{Huang2008}}\\
        $\bise$ & 10.012 & 9.84 [47] \textcolor{white}{ \cite{Nakajima1963}} & -\\
        PbTe & 6.5337 & 6.464 [49] \textcolor{white}{ \cite{Dalven1969}} & -\\
        SiGe & 5.5916 & 5.527 [51] \textcolor{white}{\cite{Steele1958}} & 5.5955 [26] \textcolor{white}{ \cite{Zhao2016}}\\
        
     \midrule
    2D Material \\
    \hdashline
        $\bite$  & 4.4162 & - & 4.38  [50] \textcolor{white}{ \cite{Sharma2016}}\\
        $\bise$  & 4.1653 & 4.13 [11] \textcolor{white}{\cite{Sun2012}} & - \\
        PbTe  & 6.5321 & - & - \\
        SiGe & 3.9549 & - & 3.91 [20] \textcolor{white}{ \cite{Sannyal2019}}\\
    \bottomrule
    \end{tabular}}
    \label{tab.lattice}
\end{table}

\begin{table}[h]
    \centering
    \caption{Optimized chemical potentials and their corresponding carrier concentrations and ZT values}
    \scalebox{0.7}{
    \begin{tabular}{c c c c}
    \toprule
         & Maximum ZT & $ \mathlarger{\mathlarger{\mu_{opt}}}\ (eV) $ & $n\ (10^{19}\ \ cm^{-3})$\\
     \midrule
    3D Material & & &  \\
    \hdashline
        \quad $\bite$ & 0.54 & -0.084 & 2.02 (p) \\
        \quad $\bise$ & 0.18 & -0.208 & 17.6 (p) \\
        \quad PbTe & 0.49 & -0.131 & 20.1 (p) \\
        \quad SiGe & 0.09 & 0.648 & 9.40 (n)\\
    
     \midrule
    2D Material & & &  $n\ (10^{13} \ \ cm^{-2})$\\
    \hdashline
        \quad $\bite$ & 0.57 & 0.016 & 2.74 (n)\\
        \quad $\bise$ & 0.43 & 0.039 & 2.40 (n)\\
        \quad PbTe & 0.32 & 0.121 & 9.40 (p)\\
        \quad SiGe & 0.04 & -0.066 & 3.25 (p)\\
    \bottomrule
    \end{tabular}}
    \label{tab.optimized_zt}
\end{table}


\begin{thebibliography}{10}
\expandafter\ifx\csname url\endcsname\relax
  \def\url#1{{\tt #1}}\fi
\expandafter\ifx\csname urlprefix\endcsname\relax\def\urlprefix{URL }\fi
\providecommand{\eprint}[2][]{\url{#2}}

\bibitem{Hasdeo2019}
Hasdeo E~H, Krisna L~P, Hanna M~Y, Gunara B~E, Hung N~T and Nugraha A~R 2019
  {\em Journal of Applied Physics\/} {\bf 126} 1--10

\bibitem{Mahan1994}
Sofo J~O and Mahan G~D 1994 {\em Phys. Rev. B\/} {\bf 49}(7) 4565--4570

\bibitem{Chen2011}
Zhou J, Yang R, Chen G and Dresselhaus M~S 2011 {\em Phys. Rev. Lett.\/} {\bf
  107}(22) 226601

\bibitem{Hicks1993}
Hicks L~D and Dresselhaus M~S 1993 {\em Phys. Rev. B\/} {\bf 47}(19)
  12727--12731

\bibitem{Hicks1993_1D}
Hicks L~D and Dresselhaus M~S 1993 {\em Phys. Rev. B\/} {\bf 47}(24)
  16631--16634

\bibitem{Biswas2012}
Biswas K, He J, Blum I~D, Wu C~I, Hogan T~P, Seidman D~N, Dravid V~P and
  Kanatzidis M~G 2012 {\em Nature\/} {\bf 489} 414--418

\bibitem{Hochbaum2008}
Hochbaum A~I, Chen R, Delgado R~D, Liang W, Garnett E~C, Najarian M, Majumdar A
  and Yang P 2008 {\em Nature\/} {\bf 451} 163--167

\bibitem{Poudel2008}
Poudel B, Hao Q, Ma Y, Lan Y, Minnich A, Yu B, Yan X, Wang D, Muto A, Vashaee
  D, Chen X, Liu J, Dresselhaus M~S, Chen G and Ren Z 2008 {\em Science\/} {\bf
  320} 634--638

\bibitem{Hung2016}
Hung N~T, Hasdeo E~H, Nugraha A~R, Dresselhaus M~S and Saito R 2016 {\em
  Physical Review Letters\/} {\bf 117} 1--5

\bibitem{Teweldebrhan2010}
Teweldebrhan D, Goyal V and Balandin A~A 2010 {\em Nano Letters\/} {\bf 10}
  1209--1218

\bibitem{Sun2012}
Sun Y, Cheng H, Gao S, Liu Q, Sun Z, Xiao C, Wu C, Wei S and Xie Y 2012 {\em
  Journal of the American Chemical Society\/} {\bf 134} 20294--20297

\bibitem{Liu2015}
Liu J, Qian X and Fu L 2015 {\em Nano Letters\/} {\bf 15} 2657--2661

\bibitem{Jamdagni2015}
Jamdagni P, Kumar A, Thakur A, Pandey R and Ahluwalia P~K 2015 {\em Materials
  Research Express\/} {\bf 2} 16301

\bibitem{Mori2016}
Mori K, Usui H, Sakakibara H, Kuroki K, Mori K, Usui H and Sakakibara H 2016
  {\bf 042108}

\bibitem{Giannozzi2009}
Giannozzi P, Baroni S, Bonini N, Calandra M, Car R, Cavazzoni C, Ceresoli D,
  Chiarotti G~L, Cococcioni M, Dabo I, Corso A~D, de~Gironcoli S, Fabris S,
  Fratesi G, Gebauer R, Gerstmann U, Gougoussis C, Kokalj A, Lazzeri M,
  Martin-Samos L, Marzari N, Mauri F, Mazzarello R, Paolini S, Pasquarello A,
  Paulatto L, Sbraccia C, Scandolo S, Sclauzero G, Seitsonen A~P, Smogunov A,
  Umari P and Wentzcovitch R~M 2009 {\em Journal of Physics: Condensed
  Matter\/} {\bf 21} 395502

\bibitem{Kresse1999}
Kresse G and Joubert D 1999 {\em Phys. Rev. B\/} {\bf 59}(3) 1758--1775

\bibitem{Perdew1992}
Perdew J~P, Chevary J~A, Vosko S~H, Jackson K~A, Pederson M~R, Singh D~J and
  Fiolhais C 1992 {\em Phys. Rev. B\/} {\bf 46}(11) 6671--6687

\bibitem{Monkhorst1976}
Monkhorst H~J and Pack J~D 1976 {\em Phys. Rev. B\/} {\bf 13}(12) 5188--5192

\bibitem{Madsen2006}
Madsen G~K and Singh D~J 2006 {\em Computer Physics Communications\/} {\bf 175}
  67--71

\bibitem{supplementary}
 \textbf{Supplementary material} link is provided by the publisher

\bibitem{Jia2017}
Jia Y~Z, Ji W~X, Zhang C~W, Li P, Zhang S~F, Wang P~J, Li S~S and Yan S~S 2017
  {\em Physical Chemistry Chemical Physics\/} {\bf 19} 29647--29652

\bibitem{Sannyal2019}
Sannyal A, Ahn Y and Jang J 2019 {\em Computational Materials Science\/} {\bf
  165} 121--128

\bibitem{Witting2019}
Witting I~T, Chasapis T~C, Ricci F, Peters M, Heinz N~A, Hautier G and Snyder
  G~J 2019 {\em Advanced Electronic Materials\/} {\bf 5} 1--20

\bibitem{Zhou2015}
Zhou G and Wang D 2015 {\em Scientific Reports\/} {\bf 5} 1--6

\bibitem{Ryu2016}
Ryu B and Oh M~W 2016 {\em Journal of the Korean Ceramic Society\/} {\bf 53}
  273--281

\bibitem{Park2016}
Park S and Ryu B 2016 {\em Journal of the Korean Physical Society\/} {\bf 69}
  1683--1687

\bibitem{Wang2007}
Wang Y, Chen X, Cui T, Niu Y, Wang Y, Wang M, Ma Y and Zou G 2007 {\em Physical
  Review B - Condensed Matter and Materials Physics\/} {\bf 76} 1--10

\bibitem{Zhao2016}
Zhao Z~Y, Yang W and Yang P~Z 2016 {\em Chinese Physics B\/} {\bf 25}

\bibitem{Markov2019}
Markov M, Rezaei S~E, Sadeghi S~N, Esfarjani K and Zebarjadi M 2019 {\em
  Physical Review Materials\/} {\bf 3} 1--7

\bibitem{Mahan1996}
Mahan G~D and Sofo J~O 1996 {\em Proceedings of the National Academy of
  Sciences\/} {\bf 93} 7436--7439

\bibitem{Jeong2012}
Jeong C, Kim R and Lundstrom M~S 2012 {\em Journal of Applied Physics\/} {\bf
  111} 0--12

\bibitem{Nurhuda_2020}
Nurhuda M, Nugraha A~R~T, Hanna M~Y, Suprayoga E and Hasdeo E~H 2020 {\em
  Advances in Natural Sciences: Nanoscience and Nanotechnology\/} {\bf 11}
  015012

\bibitem{Xue2015}
Mi X~Y, Yu X, Yao K~L, Huang X, Yang N and LÃŒ J~T 2015 {\em Nano Letters\/}
  {\bf 15} 5229--5234

\bibitem{Ding2019}
Ding Z, An M, Mo S, Yu X, Jin Z, Liao Y, Esfarjani K, LÃŒ J~T, Shiomi J and
  Yang N 2019 {\em J. Mater. Chem. A\/} {\bf 7}(5) 2114--2121

\bibitem{Qiu2010}
Qiu B and Ruan X 2010 {\em Applied Physics Letters\/} {\bf 97} 2--4

\bibitem{Liu2017}
Liu R, Tan X, Ren G, Liu Y, Zhou Z, Liu C, Lin Y and Nan C 2017 {\em
  Crystals\/} {\bf 7}

\bibitem{Masaki2000}
Orihashi M, Noda Y, Chen L and Hirai T 2000 {\em Materials Transactions, JIM\/}
  {\bf 41} 1196--1201

\bibitem{Tayebi2012}
{Tayebi} L, {Zamanipour} Z, {Mozafari} M, {Norouzzadeh} P, {Krasinski} J~S,
  {Ede} K~F and {Vashaee} D 2012 Thermal and thermoelectric properties of
  nanostructured versus crystalline sige {\em IEEE Green Technologies
  Conference\/} (Tulsa, OK, USA) p 1--4

\bibitem{Jeon1991}
Jeon H~W, Ha H~P, Hyun D~B and Shim J~D 1991 {\em Journal of Physics and
  Chemistry of Solids\/} {\bf 52} 579--585

\bibitem{Plechacek2004}
Plech{\'{a}}{\v{c}}ek T, Navr{\'{a}}til J, Hor{\'{a}}k J and
  Lo{\v{s}}t'{\'{a}}k P 2004 {\em Philosophical Magazine\/} {\bf 84} 2217--2228

\bibitem{Kulbachinskii2012}
Kulbachinskii V~A, Kytin V~G, Kudryashov A~A and Tarasov P~M 2012 {\em Journal
  of Solid State Chemistry\/} {\bf 193} 47--52

\bibitem{Hong2015}
Hong M, Chen Z~G, Yang L, Han G and Zou J 2015 {\em Advanced Electronic
  Materials\/} {\bf 1} 1--9

\bibitem{Hor2009}
Hor Y~S, Richardella A, Roushan P, Xia Y, Checkelsky J~G, Yazdani A, Hasan M~Z,
  Ong N~P and Cava R~J 2009 {\em Physical Review B - Condensed Matter and
  Materials Physics\/} {\bf 79} 2--6

\bibitem{McGuire2008}
McGuire M~A, Malik A~S and DiSalvo F~J 2008 {\em Journal of Alloys and
  Compounds\/} {\bf 460} 8--12

\bibitem{Basu2013}
Basu R, Bhattacharya S, Bhatt R, Singh A, Aswal D~K and Gupta S~K 2013 {\em
  Journal of Electronic Materials\/} {\bf 42} 2292--2296

\bibitem{Pei2012}
Pei Y~L and Liu Y 2012 {\em Journal of Alloys and Compounds\/} {\bf 514} 40--44

\bibitem{Nozariasbmarz2016}
Nozariasbmarz A, Roy P, Zamanipour Z, Dycus J~H, Cabral M~J, LeBeau J~M,
  Krasinski J~S and Vashaee D 2016 {\em APL Materials\/} {\bf 4}

\bibitem{Wang2008}
Wang X~W, Lee H, Lan Y~C, Zhu G~H, Joshi G, Wang D~Z, Yang J, Muto A~J, Tang
  M~Y, Klatsky J, Song S, Dresselhaus M~S, Chen G and Ren Z~F 2008 {\em Applied
  Physics Letters\/} {\bf 93} 21--24

\bibitem{Bathula2012}
Bathula S, Jayasimhadri M, Singh N, Srivastava A~K, Pulikkotil J, Dhar A and
  Budhani R~C 2012 {\em Applied Physics Letters\/} {\bf 101}

\bibitem{Janicek2009}
Janíček, P. and Drašar, Č. and Beneš, L. and Lošťák, P 2009 {\em Crystal Research and
  Technology\/} {\bf 44} 505--510

\bibitem{Scheidemantel2003}
Scheidemantel J, Ambrosch-Draxl C, Thonhauser T, Badding V and Sofo O 2003 {\em
  Physical Review B - Condensed Matter and Materials Physics\/} {\bf 68} 1--6

\bibitem{Goldsmid_1958}
Goldsmid H~J, Sheard A~R and Wright D~A 1958 {\em British Journal of Applied
  Physics\/} {\bf 9} 365--370

\bibitem{Hunter2007}
Hunter J~D 2007 {\em Computing in Science \& Engineering\/} {\bf 9} 90--95

\bibitem{Nakajima1963}
Nakajima S 1963 {\em Journal of Physics and Chemistry of Solids\/} {\bf 24}
  479--485

\bibitem{Huang2008}
Huang B~L and Kaviany M 2008 {\em Physical Review B - Condensed Matter and
  Materials Physics\/} {\bf 77} 1--19

\bibitem{Dalven1969}
Dalven R 1969 {\em Infrared Physics\/} {\bf 9} 141--184

\bibitem{Sharma2016}
Sharma S and Schwingenschl{\"{o}}gl U 2016 {\em ACS Energy Letters\/} {\bf 1}
  875--879

\bibitem{Steele1958}
Steele M~C and Rosi F~D 1958 {\em Journal of Applied Physics\/} {\bf 29}
  1517--1520

\end{thebibliography}
\end{document}